\documentclass{iau}
\usepackage{graphicx}

\title[225 GHz Opacity at 2 Arctic Sites] 
{225 GHz Atmospheric Opacity Measurements from Two Arctic Sites}

\author[S. Matsushita et al.]   
{S. Matsushita$^1$, Ming-Tang Chen$^1$, P. Martin-Cocher$^1$,
 K. Asada$^1$, C.-P. Chen$^1$, M. Inoue$^1$, S. Paine$^2$,
 D. Turner$^3$, \and E. Steinbring$^4$}

\affiliation{$^1$Institute of Astronomy and Astrophysics,
	Academia Sinica, \\ P.O.Box 23-141, Taipei 10617, Taiwan, R.O.C. \\
	email: {\tt satoki@asiaa.sinica.edu.tw} \\[\affilskip]
$^2$Smithsonian Astrophysical Observatory, \\
	160 Concord Ave., Cambridge, MA 02138, USA\\[\affilskip]
$^3$National Severe Storms Laboratory, \\
	120 David L. Boren Boulevard, Norman, OK, 73072, USA\\[\affilskip]
$^4$National Research Council, Herzberg Inst of Astrophysics, \\
	5071 W Saanich Rd, CA Victoria BC V9E 2E7, Canada}

\pubyear{2012}
\volume{288}  
\pagerange{119--126}
\jname{Astrophysics from Antarctica}
\editors{M.G. Burton, X. Cui \& N.F.H. Tothill, eds.}
\begin{document}

\maketitle

\begin{abstract}
We report the latest results of 225 GHz atmospheric opacity
measurements from two arctic sites; one on high coastal terrain near
the Eureka weather station, on Ellesmere Island, Canada, and the
other at the Summit Station near the peak of the Greenland icecap.
This is a campaign to search for a site to deploy a new telescope for
submillimeter Very Long Baseline Interferometry and THz astronomy in
the northern hemisphere.
Since 2011, we have obtained 3 months of winter data near Eureka, and
about one year of data at the Summit Station.
The results indicate that these sites offer a highly transparent
atmosphere for observations in submillimeter wavelengths.
The Summit Station is particularly excellent, and its zenith opacity
at 225 GHz is statistically similar to the Atacama Large
Milllimeter/submillimeter Array in Chile.  In winter, the opacity at
the Summit Station is even comparable to that observed at the South
Pole.
\end{abstract}

\firstsection 
\section{Introduction}

The success of \cite[Doeleman et al. (2008)]{doe08} in obtaining a
scatter-free size estimate of submillimeter emission in Sagittarius
A$^*$ (Sgr A$^*$) using Very Long Baseline Interferometry (VLBI)
promises a new window for direct imaging of supermassive black holes
(SMBHs).
Although Sgr A$^*$ is the nearest known and biggest (in apparent
size), its mass is relatively small among SMBH.
This means a short timescale of variability, leading to undesirable
smoothing of that signal during integration.
On the other hand, the second largest source in apparent size is the
SMBH in M87 (Virgo A), which has a large mass.
A further scientifically interesting point is that M87 has a strong
jet activity.
But to image the SMBH in M87 with submm-VLBI requires a longer
baseline in the northern hemisphere than currently available (see
\cite[Inoue et al. 2012]{ino12} in this proceeding) and so we began
a search for a new submm telescope site.

\firstsection

\section{Site Selection}

For the site selection, we set criteria as follows:
(1) Annual precipitable water vapor (PWV) of less than 3 mm for good
	submm opacity.
(2) Longest-possible baseline with existing telescopes, for best
	imaging resolution.
(3) Observable sky together with the Atacama Large
	Millimeter/submillimeter Array (ALMA) to achieve highest possible
	sensitivity.
(4) Accessibility to the site.

Based on the first two criteria, there are three potential broad
regions of interest; western China and Tibet, the highest mountains
of southern Alaska, or the High Arctic polar desert, including
northern Canada and Greenland.
The Western China and Tibet region does not have common sky with
ALMA, so it does not meet criterion (3), and the tallest peaks in
Alaska (e.g., Mount McKinley) are excluded due to criterion (4).
The Eureka research base on Ellesmere Island, Canada, and Summit on
the Greenland icecap meet all four criteria, and we considered these
two sites for further study.

\firstsection

\section{225 GHz Tipping Radiometer}

For the site survey, we purchased a 225 GHz tipping radiometer from
the Radiometer Physics GmbH.
The reason for the choice of this frequency is that there are many
site survey results from all over the world, including the summit of
Mauna Kea, the ALMA (Chajnantor) site, and South Pole.

For the opacity measurements, we use the tipping method; we observe
five angles ($90^{\circ}$, $42^{\circ}$, $30^{\circ}$, $24^{\circ}$,
and $19.2^{\circ}$, which corresponds to $\sec(z)$ of 1.0, 1.5, 2.0,
2.5, and 3.0) with 4 second integration at each angle, for a duration
of 75 second per tipping measurement.
And for each, the opacity is derived from the instrument output
voltage as a function of zenith angle.
Measurements were obtained every 10 minute.

We first tested at the Academia Sinica, Institute of Astronomy and
Astrophysics (ASIAA) in Taipei, Taiwan, and then repeated on Mauna
Kea, Hawaii to check the consistency of our measurements with the
225 GHz tipping radiometer at the Caltech Submillimeter Observatory
(CSO).
Simultaneous opacity measurements were performed between 31 December
2010 and 11 January 2011, and are consistent with each other (linear
regression coefficient = 1.04).

\firstsection

\section{Results}

Identical observations were then carried out at the High Arctic
sites, allowing direct comparison to those worldwide.

\subsection{Eureka, Ellesmere Island, Canada}

\begin{figure}[t]
 \vspace*{-3.0 cm}
\begin{center}
 \includegraphics[width=2.15in]{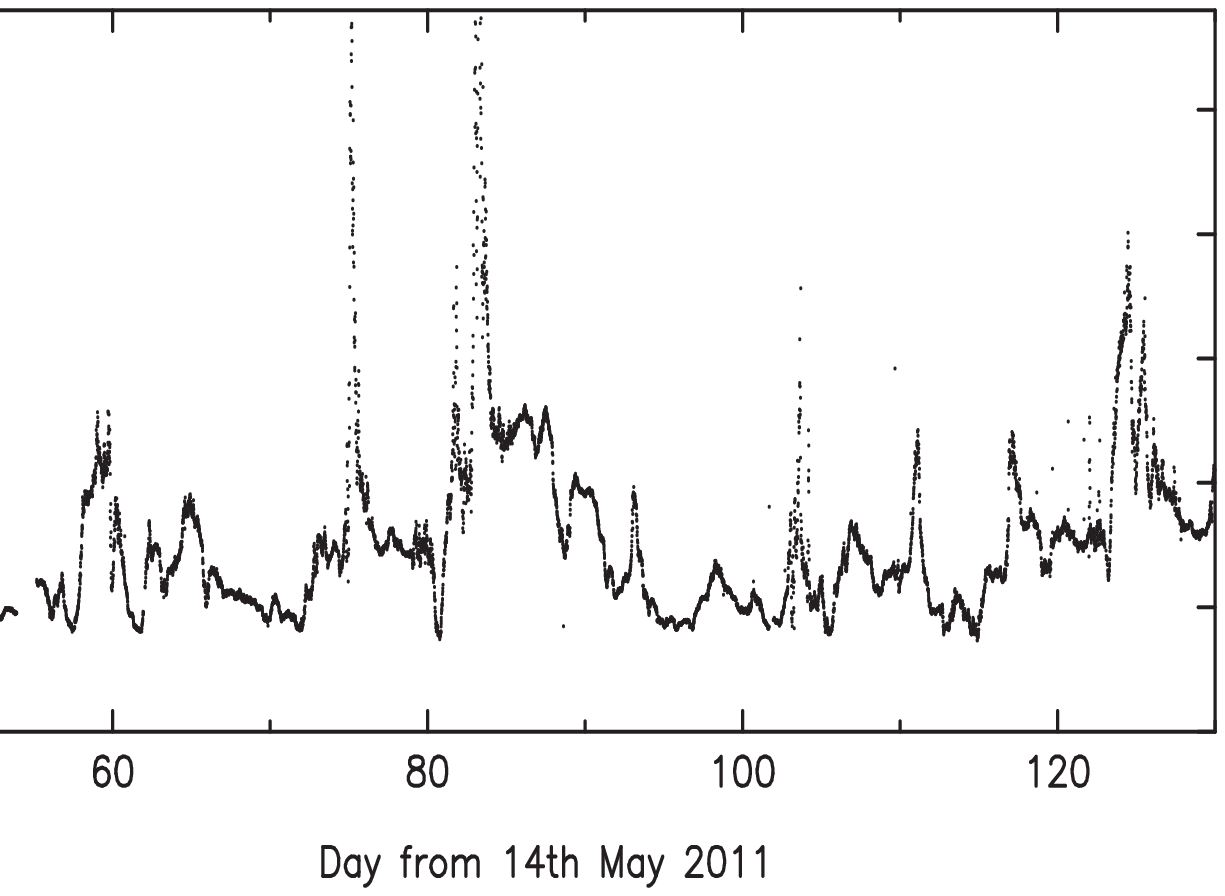}
 \hspace*{0.3 cm}
 \includegraphics[width=2.15in]{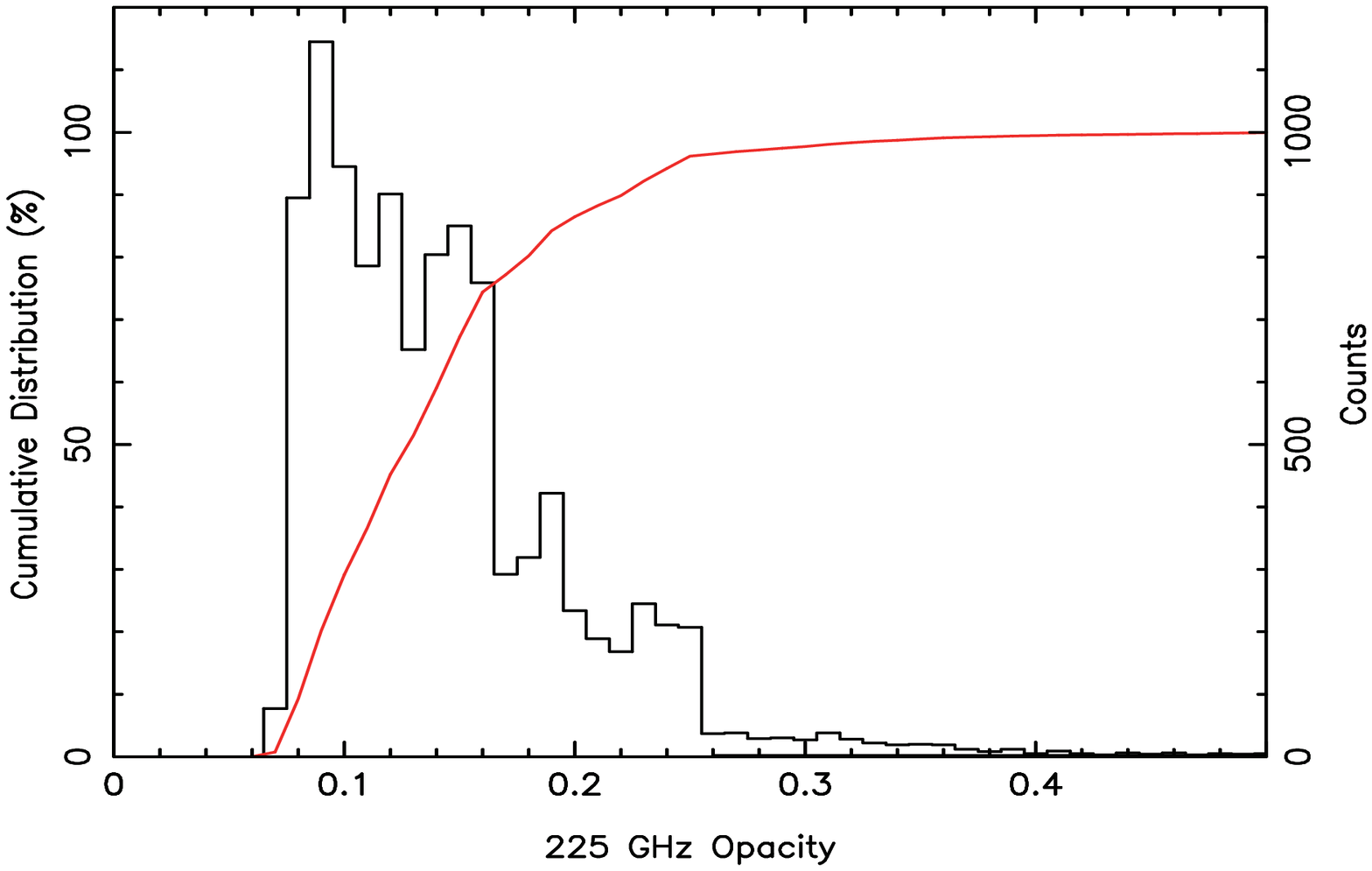}
 \vspace*{-4.0 cm}
 \caption{{\it (Left)} Time variation plot of 225 GHz opacity at the
  PEARL site.
  {\it (Right)} Histogram and cumulative plot.}
 \label{fig1}
\end{center}
\end{figure}

Eureka is a manned weather station on Ellesmere Island, Canada.
It has a 5000 foot all-season airstrip.
Yearly resupply by ship occurs in late summer.
A road allows access to the Polar Environment Atmospheric Research
Laboratory (PEARL), located on a 610 m-high ridge at N$80.05^\circ$,
W$86.42^\circ$.
Although this site is not as high as other good submm sites, it is
expected to have low opacity conditions based on low temperatures,
typically between $-20^{\circ}$C and $-40^{\circ}$C in winter
(\cite[Steinbring et al. 2010]{ste00}).
The radiometer was deployed on the rooftop observing platform of
PEARL, and measured the 225 GHz opacity between 14 February and 10
May 2011.
Tipping direction was south, providing 10,522 data points.

Fig.\,\ref{fig1} shows the time variation, histogram, and cumulative
plots.
The lowest 225 GHz opacity measured was 0.07, the 25\% quartile was
0.11, the median (50\% quartile) was 0.14, and the most frequent
opacity was 0.09 during our measurements.
These statistics indicate that submm-VLBI at this frequency is
feasible at this site.

\subsection{Summit Camp on Greenland}

Summit Camp is a research station near the peak of the Greenland ice
sheet, located at N$72.57^\circ$, W$38.46^\circ$, with an elevation
of 3200 m.
Access is by C-130 air transport in summer and by Twin Otter in
winter, or traverse from Thule.
Very good opacity conditions are expected for this site, based on low
winter temperature between $-40^{\circ}$C and $-60^{\circ}$C with the
lowest temperature of $-72^{\circ}$C
(\cite[Vaarby-Laursen 2010]{vaa10}) and the high altitude.
We put our radiometer on the roof of the Mobile Science Facility
(MSF), and started measuring the opacity on 17 August 2011 (and still
continue to do so).
Here we show data up to 31 July 2012.
Tipping directions were both south and north, that is in two
directions, and provide 36,555 data points.

\begin{figure}[t]
 \vspace*{-1.8 cm}
\begin{center}
 \includegraphics[width=2.6in]{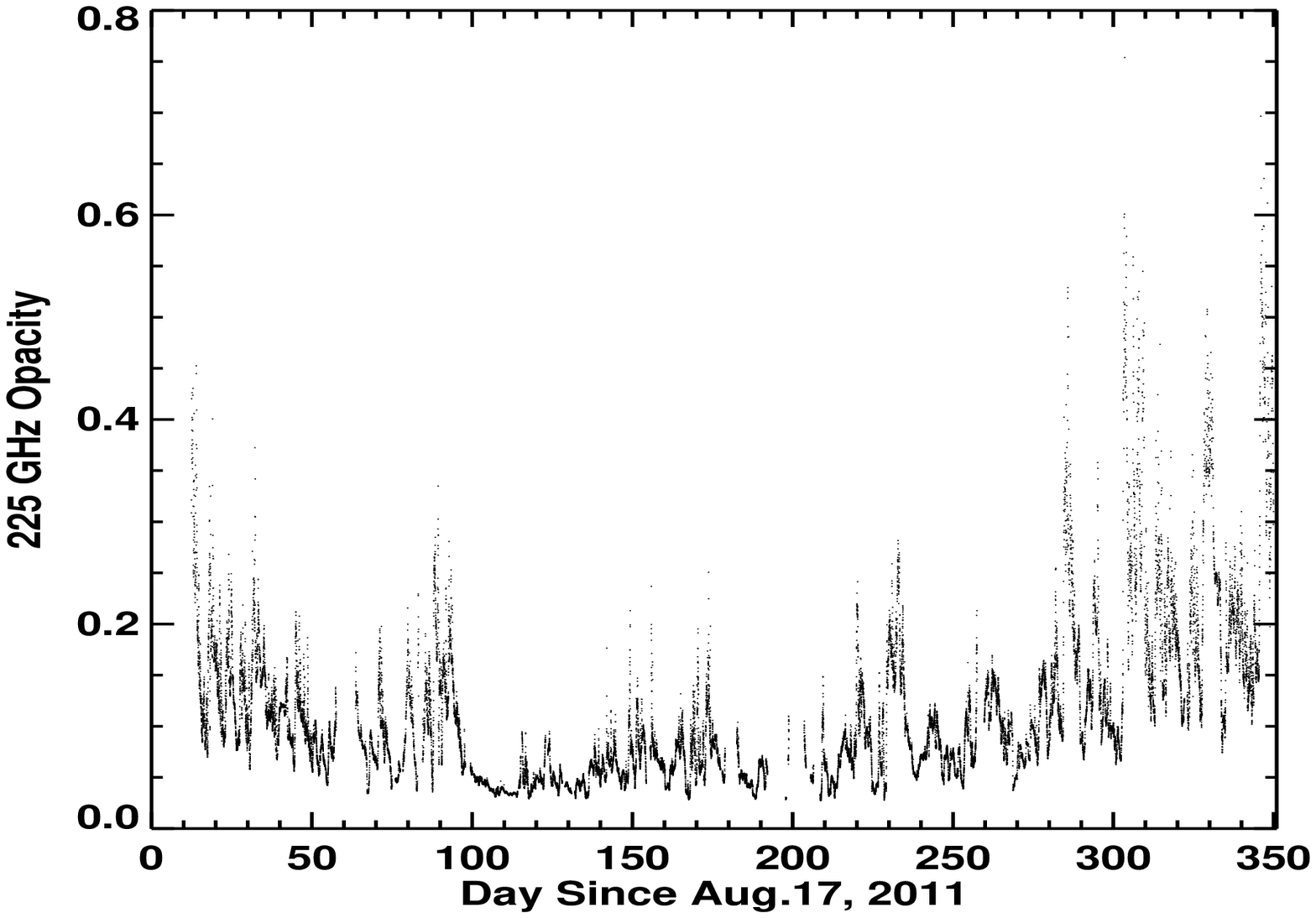}
 \hspace*{0.05 cm}
 \includegraphics[width=2.6in]{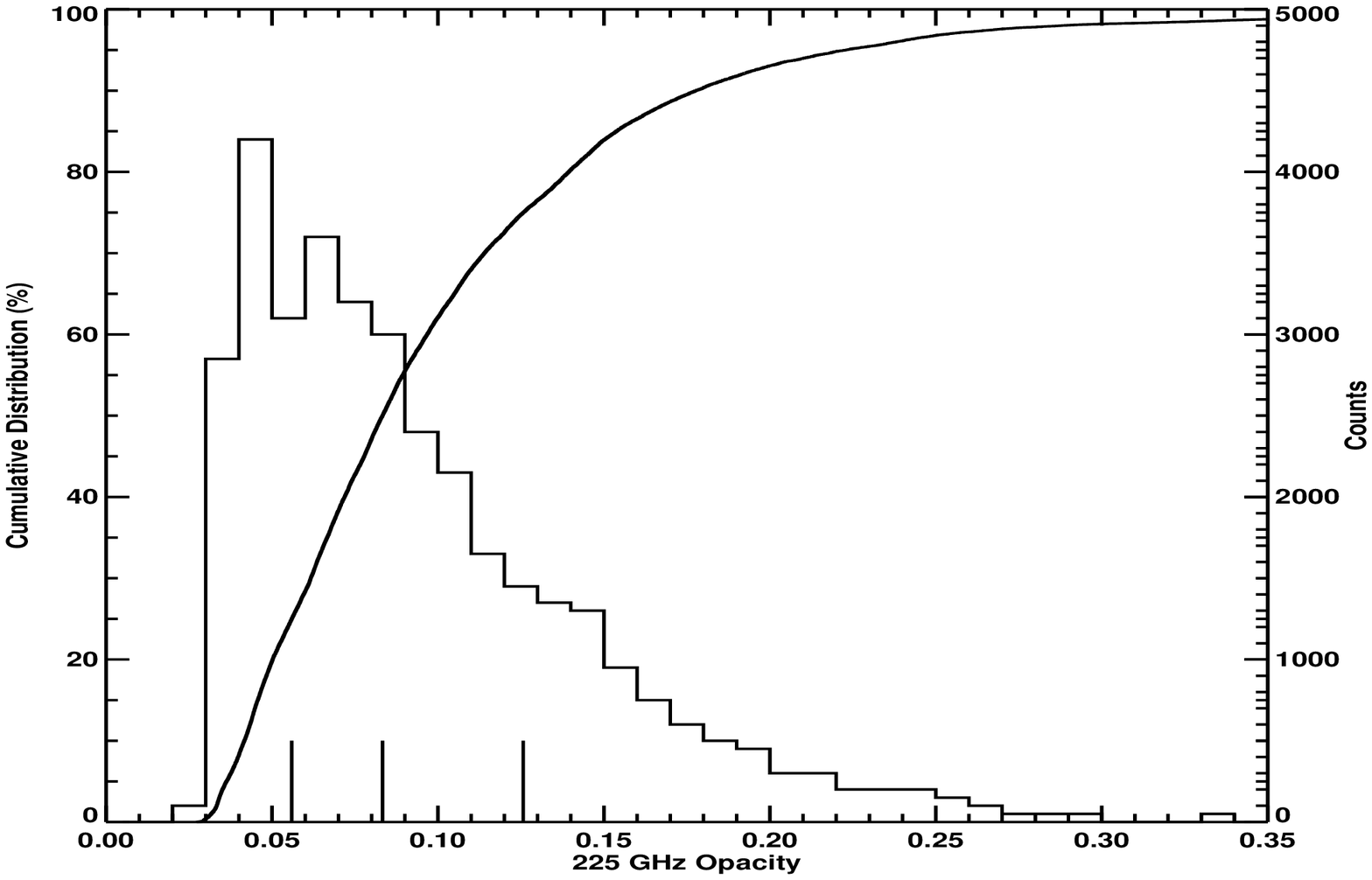}
 \caption{{\it (Left)} Time variation plot of 225 GHz opacity at the
  Summit Camp.
  {\it (Right)} Histogram and cumulative plot.}
 \label{fig2}
\end{center}
\end{figure}

Fig.\,\ref{fig2} shows the time variation, histogram, and cumulative
plots.
The lowest 225 GHz opacity measured was 0.027, the 25\% quartile was
0.056, the median was 0.083, and the most frequent opacity was 0.04
during our measurements.
These statistics are excellent, strong indication that submm-VLBI at
these frequencies or even higher are feasible at this site.
The THz astronomy is also worth to consider.

\subsection{Comparison with Other Sites}

To illustrate how a High Arctic submm site can compare to the best
worldwide, we compared Summit Camp statistics with those of the
ALMA site (elevation = 5050 m) and South Pole (2800 m).
The opacity data for the ALMA site has been taken from
\cite[Radford \& Chamberlin (2000)]{rad00} and
\cite[Radford (2011)]{rad11}, which has been measured between April
1995 and April 2006, and that for South Pole from
\cite[Chamberlin \& Bally (1994)]{cha94} and
\cite[Chamberlin \& Bally (1995)]{cha95}, which has been measured
between January and December 1992.

\begin{figure}[t]
\begin{center}
 \includegraphics[width=5.3in]{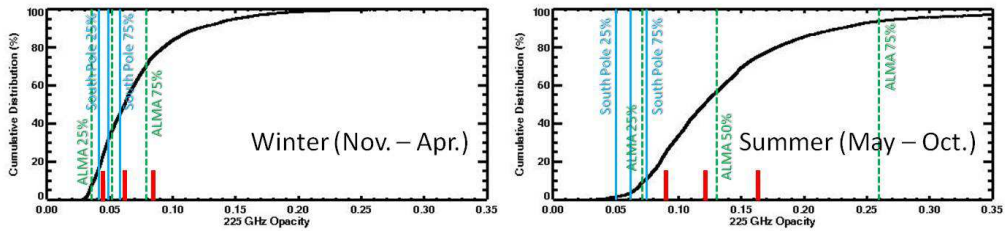}
 \caption{Opacity quartiles comparison between the Summit Camp
  (histogram and cumulative plots with quartiles as short and thick
  solid lines), the ALMA site (dashed vertical lines), and South
  Pole (thin solid vertical lines).
  Left plots are winter season, and right summer.}
 \label{fig3}
\end{center}
\end{figure}

Winter in the northern hemisphere is defined as between the beginning
of November and the end of April, with summer May through October.
The opposite is taken to be the case for the southern hemisphere;
winter May through October, summer November through April.
We put the quartiles for the three sites on the cumulative plots of
the Summit Camp in Fig.\,\ref{fig3}.

In winter season, 225 GHz opacities of all three sites are less than
0.1 even at 75\% quartile.
There is small difference between the sites, but the opacity
statistic of the Summit Camp is rather similar to that of the ALMA
site.
The opacity statistics of South Pole do not vary much winter to
summer, but the ALMA site shows much greater variation.
Summit Camp falls somewhere in between.
It is worth noting, however, that 2011/12 represents a record warm
period for Greenland, which may have affected opacity statistics.
Long-term monitoring is needed to carefully judge the quality of
Summit Camp as a submm site.

\firstsection

\section{Summary}

We have presented a program of site testing for a submm telescope
site in the northern hemisphere.
Both Eureka, Ellesmere Island, and Summit Camp, Greenland offer the
potential for new submm VLBI observations.
Based on the best 225 GHz opacity measurements, we selected Summit
Camp, and current efforts are underway on retrofitting an antenna for
the extremes of the site.
The opacity measurements are still ongoing to collect long-term
opacity variation data.
It will reveal whether the record warmth in Greenland has affected
the opacity statistics or not, and the true fraction of time that
Summit Camp reaches the quality of South Pole.
In addition, atmospheric characteristics using various
instrumentations (cloud radars and lidars, radiosondes, microwave
radiometers, precipitation measurements) are ongoing at Summit Camp
by atmospheric researchers (e.g., \cite[Shupe et al. 2012]{shu12}).
We are closely collaborating with them to estimate the atmospheric
conditions more accurately, and to construct accurate atmospheric
models for the coming submm/THz astronomy at this site.

%
%
%
%
%
%
%


\begin{thebibliography}{}

\bibitem[Chamberlin \& Bally (1994)]{cha94}
{Chamberlin, R.A. \& Bally, J.} 1994, \textit{App. Opt.} 33, 1095

\bibitem[Chamberlin \& Bally (1995)]{cha95}
{Chamberlin, R.A. \& Bally, J.} 1995,
\textit{Int. J. IR MM Waves} 16, 907

\bibitem[Doeleman et al. (2008)]{doe08}
{Doeleman, S.S., et al.} 2008, \textit{Nature} 455, 78

\bibitem[Inoue et al. (2012)]{ino12}
{Inoue, M., et al.} 2012, this proceeding

\bibitem[Radford (2011)]{rad11}
{Radford, S.J.E.} 2011, \textit{RevMexAA (SC)} 41, 87

\bibitem[Radford \& Chamberlin (2000)]{rad00}
{Radford, S.J.E. \& Chamberlin, R.A.} 2000, \textit{ALMA Memo} 334

\bibitem[Shupe et al. (2012)]{shu12}
{Shupe, M.D., et al.} 2012, \textit{Bull. Amer. Meteo. Soc.} in press

\bibitem[Steinbring et al. (2010)]{ste10}
{Steinbring, E., Carlberg, R., Croll, B., Fahlman, G., Hickson, P.,
 Ivanescu, L., Leckie, B., Pfrommer, T. \& Schoeck, M.} 2010,
\textit{PASP} 122, 1092

\bibitem[Vaarby-Laursen (2010)]{vaa10}
Vaarby-Laursen, E., \textit{DMI Tech. Rep.} 10-09

\end{thebibliography}
\end{document}